\documentclass{article}
\usepackage{color}
\usepackage{booktabs}
\usepackage{array}
\usepackage{tabularray}
\usepackage{pifont}
\usepackage{amsmath,amssymb,amsfonts}
\usepackage{algorithmic}
\usepackage{graphicx}
\usepackage{multirow}
\usepackage{svg}
\usepackage{textcomp}
\usepackage{xcolor}
\usepackage{diagbox}
\usepackage{adjustbox}
\usepackage{hyperref}
\usepackage{ctable}
\usepackage[affil-it]{authblk}
\hypersetup{
    colorlinks,
    citecolor=black,
    filecolor=black,
    linkcolor=black,
    urlcolor=black
}
\usepackage[square,numbers, sort&compress]{natbib}

\begin{document}

\title{Review, Definition and Challenges of Electrical Energy Hubs}

\author[1,2]{Giacomo Bastianel}
\author[1,2]{Jan Kircheis}
\author[1,2]{Merijn {Van Deyck}}
\author[1,2]{Dongyeong Lee}
\author[1,2]{Geraint Chaffey}
\author[1,2]{Marta Vanin}
\author[1,2]{Hakan Ergun}
\author[1,2]{Jef Beerten}
\author[1,2]{Dirk {Van Hertem}}

\affil[1]{{Department of Electrical Engineering, KU Leuven, Belgium}}
\affil[2]{{Etch-EnergyVille, Belgium}}

\date{}
\maketitle

\begin{abstract}
To transition towards a carbon-neutral power system, considerable amounts of renewable energy generation capacity are being installed in the North Sea area.
Consequently, projects aggregating many gigawatts of power generation capacity and transmitting renewable energy to the main load centers are being developed. Given the electrical challenges arising from having bulk power capacity in a compact geographical area with several connections to the main grid, and a lack of a robust definition identifying the type of system under study, this paper proposes a general technical definition of such projects introducing the term \textit{Electrical Energy Hub (EEH)}.
The concept, purpose, and functionalities of EEHs are introduced in the text, emphasizing the importance of a clear technical definition for future planning procedures, grid codes, regulations, and support schemes for EEHs and multiterminal HVDC (MTDC) grids in general. Furthermore, the unique electrical challenges associated with integrating EEHs into the power system are discussed. Three research areas of concern are identified, namely control, planning, and protection. Through this analysis, insights are provided into the effective implementation of multi-GW scale EEH projects and their integration into the power grid through multiple interconnections.
Finally, a list of ongoing and planned grid development projects is evaluated to assess whether they fall within the EEH category. 
\end{abstract}


\section{Introduction} \label{sec:introduction}
 Climate change and the constant increase in the average global temperature compared to pre-industrial levels have triggered a rapid transition towards net-zero power systems. To achieve a decrease in $CO_{2}$ emissions, considerable amounts of renewable energy sources (RES) are installed, with a cumulative worldwide capacity of 1185~GW of solar PV~\cite{PV_report} and 906~GW of wind power~\cite{Wind_report} in 2023. Among the installed wind power capacity, offshore wind grew from 2.1~GW in 2009 to 72.7~GW in 2023. In addition, 380~GW of offshore wind generation capacity is expected to be installed worldwide in the next decade~\cite{Wind_report}. In this regard, the European Union (EU) foresees 300~GW of installed offshore wind capacity in its territory by 2050~\cite{RenewableEnergyDirective}. 
 As such, offshore wind plays a key role in the EU's plans to reduce the $CO_{2}$ emissions by 2030 with at least 55\% compared to 1990 levels and to make the continent carbon-neutral by 2050~\cite{fitfor55}. In particular, the governments of Belgium, Germany, the Netherlands, Denmark, France, Ireland, Luxembourg, Norway, and the United Kingdom have agreed on reaching at least 300~GW of offshore wind capacity installed in the North Sea by 2050 with the Ostend Declaration~\cite{Ostend_declaration}. So far, offshore wind power has mainly been transmitted from wind farms directly to the onshore grid, either in AC~\cite{Elia_MOG_1} or high-voltage DC (HVDC)~\cite{BorWin1}. With many wind farms installed in the North Sea area, scientific literature has proposed creating offshore grids to connect and deliver RES power to different points in onshore power systems more than a decade ago~\cite{Dirk2010DCgrids}. Recently, several institutions such as the International Energy Agency~\cite{IEA_grid}, the European Commision~\cite{Grid_action_plan}, ENTSO-E~\cite{ONDP} and the Belgian (Elia~\cite{Eliaenergyisland}), Danish (Energinet~\cite{Bornholm,Danish_Energy_Island_NS}) and Dutch-German (TenneT~\cite{Tennet_2GW}) transmission system operators have remarked the importance of offshore grids and, in general, of transmission infrastructure investments to decarbonize our society. Such offshore grids shall be predominantly based on HVDC cables, which have proven to be the most cost-effective solution for transmitting electricity over long distances~\cite{bookDirk}. Furthermore, the power generated by offshore RES can be collected in electrical nodes and transmitted to the onshore grid through interconnected HVDC links, creating a multiterminal HVDC (MTDC) grid. These grids facilitate power exchange between different control areas without congesting the existing AC grid and will likely become considerably more common in the future~\cite{ONDP,Caith_Moray}. When combining this interconnection aspect with multi-GW generation in a single unit such as in recent ``\textit{energy island}'' projects~\cite{Eliaenergyisland,Danish_Energy_Island_NS,Bornholm}, several technical challenges emerge. As there is no clear definition of these new projects, nor an in-depth description of the technical challenges related to them, this paper proposes the term \textit{electrical energy hubs} to describe such projects.

The rest of this paper is organized as follows. Section~\ref{sec:definition} includes a definition for the proposed term \textit{electrical energy hubs}, their functionalities, and the motivation behind this work. Section~\ref{sec:literature_review} positions the term in the literature, emphasizing the need for a clear definition, while Section~\ref{sec:challenges} discusses the technical challenges related to the previously listed functionalities and to which research domain they belong among control, planning, and protection. Finally, Section~\ref{sec:classification} classifies some ongoing and planned grid infrastructure projects, and Section~\ref{sec:Conclusion} concludes the paper.

\section{Motivation and definition of electrical energy hubs}\label{sec:definition}

With an increase in the distance of offshore wind farms from shore, the power transmission projects in the North Sea area have shifted from Point-to-Point (PtP) AC~\cite{Elia_MOG_1,DolWin2,DolWin3,HollandseKust} to PtP HVDC~\cite{BorWin1,BorWin2,BorWin3,DolWin6,Hornsea1} links. As HVDC transmission systems are cheaper than AC technologies over long distances~\cite{bookDirk,Stephen_powertech}, and guarantee enhanced power flow controllability and frequency support between asynchronous zones~\cite{Dirk2010DCgrids}, they are currently the preferred option for transmission grid expansion projects related to offshore wind. Moreover, several options for future development of offshore HVDC grids are being proposed, ranging from a central hub with interconnectors~\cite{Meijden_Tennet}, to several hub-and-spoke combinations with integrated hydrogen production~\cite{NSWPH}, standardized HVDC PtP links~\cite{Tennet_2GW} or MTDC/meshed offshore HVDC grids~\cite{Dirk2010DCgrids,PROMOTioN_final_report,Shetland}. The last two options coexist in the ENTSO-E Offshore Network Development Plan~\cite{ONDP}, where a combination of offshore ``\textit{hybrid transmission corridors}'' and PtP links is foreseen for the European future grid. In this regard, Elia~\cite{Eliaenergyisland} and Energinet~\cite{Bornholm,Danish_Energy_Island_NS} proposed to combine the aggregation of several GW of offshore wind power and its transmission to several European countries via HVDC interconnections in one unit, creating offshore MTDC grids. These projects follow the concept proposed in~\cite{Meijden_Tennet}. While there seems to be a general understanding of the main principles underlying the goals of these projects~\cite{MarsMission,Eliaenergyisland,Bornholm}, there is currently no consistent terminology to address them from a technical perspective. Therefore, in this paper, the following definition of the term \textit{electrical energy hub} is proposed:\\

\textit{An Electrical Energy Hub (EEH) is a multifunctional node in a power system, managed separately from the main existing control areas, which aggregates local GW-size electrical generation capacity. It is characterized by a high density of electrical equipment and has multiple interconnections to other control areas.}\\

This definition helps to classify EEH projects when clear distinctions are needed, e.g. when defining future planning procedures, grid codes, regulations, and support schemes for EEHs and MTDC grids in general. The development of EEHs implies considerable investments into technologies for which today there is no economy of scale, e.g. HVDC substations, HVDC circuit breakers (DCCB), and HVDC interconnections. A clear definition of the concept could help to accelerate permitting and funding procedures of future projects if their benefit to socio-economic welfare can be quantified and demonstrated. For example, the Projects of Common and Mutual Interest~\cite{PCI} from the European Commission are evaluated with a series of economical key performance indicators~\cite{ENTSO_E_CBA} and benefit from accelerated permitting procedures and funding. 

In addition, several EEH key \textit{functionalities} are mentioned in the proposed definition and discussed hereafter. Though these functionalities are not necessarily unique to EEHs, their combination creates new challenges for the planning, control, and protection of EEH projects. Note that throughout the paper, the term ``\textit{node}'' used in the definition stands for a substation where multiple network elements are connected.

\begin{figure}[h]
    \centering
    \includegraphics[width=0.7\linewidth]{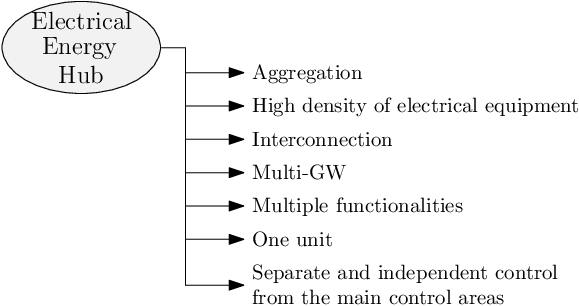}
    \caption{Functionalities characterizing electrical energy hub projects.}
    \label{fig:functional_aspects}
\end{figure}

The two main purposes of EEHs are \textit{interconnection}\footnotemark{} and \textit{aggregation}. \footnotetext{The ``interconnection'' and ``control area'' terms used in this paper refer to the definitions from the UCTE glossary~\cite{Glossary}.} Aggregation refers to collecting electrical power from several (renewable) energy sources and conditioning it for efficient transmission, e.g. transferring voltage from 66 kV AC of the wind turbines to $\pm$525 kV HVDC. Additionally, EEHs are \textit{multi-GW} electrical substations \textit{separate from the main control areas}, providing \textit{multiple interconnections} to them. \textit{Aggregation} and \textit{interconnection} are combined in \textit{one unit}. Finally, integrating \textit{multiple functionalities} in a small geographical area implies a \textit{high density of electrical equipment}, which lead to additional control and protection challenges. All of the identified functionalities are summarized in Figure~\ref{fig:functional_aspects}.

Note that while the concepts described in the paper should remain applicable to any EEH, the following sections focus on EEHs based on aggregation of offshore wind energy and interconnection of different control areas through HVDC interconnectors, as proposed by the recent projects in the North Sea~\cite{Eliaenergyisland,Danish_Energy_Island_NS,Bornholm}.

\section{Positioning of electrical energy hubs in literature} \label{sec:literature_review}

Van der Meijden~\cite{Meijden_Tennet} first proposed the concept of offshore grid nodes combining large-scale offshore wind generation capacity and interconnections to several European countries. Nevertheless, until now, no comprehensive technical definition of this concept has been proposed in the literature. Therefore, this section positions the proposed definition of EEHs in the literature, relating it to existing relevant terms for similar grid expansion projects.

The term \textit{Energy Hub} (EH) is often used to refer to planned offshore grid projects with interconnectors in the North and Baltic Sea \cite{Eliaenergyisland,Danish_Energy_Island_NS,Bornholm}. The EH concept was initially proposed for multi-energy systems by Geidl et al.~\cite{geidl2006energy}. The authors define it as ``\textit{a unit where multiple energy carriers can be converted, conditioned and stored}'' \cite{geidl2006energy}. According to this definition, EHs can be seen as units that combine various energy carriers to reduce costs and emissions and boost the reliability of supply. Although EHs are commonly linked with electricity transmission projects in the literature, the main applications include industrial plants, buildings, and ships, among others. Given the multiple applications being mentioned, specific functionalities and challenges for power grid expansion projects cannot be derived. In addition, some offshore projects with storage technologies such as hydrogen~\cite{LUTH_H2} are classified as EHs. In these cases, the projects have a broader scope than the proposed definition of EEHs, which is purely based on a power system operation perspective. 

Recently, the term \textit{Offshore Energy Hub} (OEH) has been introduced by Lüth et. al.~\cite{LUTH_2024}. It is defined as an offshore renewable-based combination of assets, where at least two services among electricity generation, interconnection, and offshore storage are provided. Although this definition gives a techno-environmental description of planned offshore grid expansion projects in the North- and Baltic Seas, it does not mention their relevant technical functionalities. As a consequence, from the definition, it is unclear whether an OEH is a single node in a grid or an entire grid itself. Drawing from the general concept from~\cite{Meijden_Tennet} and the envisioned projects for the near future~\cite{Eliaenergyisland,Danish_Energy_Island_NS,Bornholm}, it is apparent that these installations will constitute offshore nodes interconnected with multiple control areas. Even if the distinction between grid and node might not be relevant to the economic or societal context of OEHs, it is highly relevant to the electrical system design and operation of the planned projects in the North- and Baltic Sea.

Moreover, the term \textit{Energy Island} (EI) introduced by Cutululis et al. in ~\cite{MarsMission} is often used to describe the previously mentioned offshore grid projects. In~\cite{MarsMission}, EIs are defined as hubs that connect offshore wind power plants to the power system, utilizing HVDC converters from different manufacturers. In contrast to~\cite{LUTH_2024}, the definition in~\cite{MarsMission} does not mention interconnections, although the multi-purpose utilization of power lines is a key feature of the original concept proposed in~\cite{Meijden_Tennet}. In addition, the EI definition overlooks key functionalities of offshore grid projects such as~\cite{Eliaenergyisland,Danish_Energy_Island_NS,Bornholm}, e.g. the significantly higher generation capacity compared to already existing offshore installations~\cite{Hornsea1,HollandseKust}.

Overall, the existing terminology fails to adequately capture the essential technical and functional details of the envisioned offshore grid projects in the North- and Baltic Sea~\cite{Eliaenergyisland,Danish_Energy_Island_NS,Bornholm}. While concepts like EHs, OEHs, and EIs have been discussed in the literature, there is i) no consistent definition for projects such as~\cite{Eliaenergyisland,Danish_Energy_Island_NS,Bornholm} and ii) the existing definitions are often too broad. Additionally, they do not comment on technical challenges arising from EEHs. Yet, these aspects are crucial for the electrical system design and operation of EEH projects. 






\section{Technical challenges of electrical energy hubs}\label{sec:challenges}
Several technical challenges emerge from EEH projects. They are each discussed separately in this section.  In addition, Table~\ref{tab:challenges} relates them to the EEH functionalities defined previously in Figure~\ref{fig:functional_aspects}, and to the specific research areas identified as control, planning, and protection. 
\\

\textbf{\emph{Grid congestion:}} The aggregation of multi-GW power capacity and interconnections to existing control areas implies high power transmission between EEHs and existing grids. These power flows might create \textit{physical congestion} in the power grid, which is defined as ``\textit{any network situation where forecasted or realized power flows violate the thermal limits of the elements of the grid and voltage stability or the angle stability limits of the power system}''~\cite{capacity_and_congestion}. If physical congestion is recurrent on a given part of a power system, it becomes structural, reoccurring frequently under normal power system conditions~\cite{capacity_and_congestion}. Therefore, congestion might limit the maximum transferrable power from EEHs, requiring curtailment of the aggregated energy generation. Consequently, assessing the hosting capacity of the substations where the interconnections from the EEH are linked is paramount to guarantee that the existing control areas can physically transmit the power flows while avoiding the congestion of parts of the power grid~\cite{Hakan_powertech}. 

Furthermore, investments in new transmission lines or power-flow control devices such as phase-shifting transformers and switching units increase the available transmission capacity in a given power system~\cite{PSTs_placement}. Such switching units have the potential to be used flexibly to adapt the EEH topology and control. As a result, the aggregated energy generation curtailment can be minimized, bringing additional benefits to the overall socioeconomic welfare.

Switching unit actions to optimize the grid topology are being explored for existing AC grids~\cite{Morsy2022,Hinneck,Heidarifar2016,nrao} and for future MTDC grids including EEHs~\cite{Jack}. Such actions are expected to offer a cost-efficient methodology for balancing the RES and load variability and aim to increase the control areas' capacity to host the multi-GW power of EEHs. Currently, congestion is relieved mainly through redispatch, defined as ``\textit{altering the generation and/or load pattern to change physical flows in the transmission system}''~\cite{Commission_regulation}. Given the high cost for redispatch~\cite{ENTSO_E_Congestion}, the topological actions used in~\cite{nrao,Jack} should be implemented as the preferred option to relieve grid congestion while reducing the total grid operational costs.\\

\textbf{\emph{Limit maximum loss of infeed:}} Electrical energy hubs are characterized by a multi-GW power transfer capacity. Therefore, outages might cause large power deviations in the connected grids. As a consequence, a protection system is required to prevent faults from causing a loss of infeed (LoI) above the limit set by the dimensioning incident of the main control areas, e.g. 3~GW in the continental European synchronous area and 1.2~GW in the Nordic power system~\cite{EC_grid_code}. 

As the currently planned EEHs in Europe are based on HVDC interconnections~\cite{Elia2023Website,Danish_Energy_Island_NS}, the cost of the protection system is not negligible. HVDC protection equipment is expensive compared to AC equipment and should therefore be considered in the early grid planning stages to avoid excessive costs and over-design~\cite{Dirk2010DCgrids}. Nevertheless, it should be avoided that the entire power flow through multi-GW EEHs is interrupted even for a limited time (e.g. tens of ms)~\cite{DCcont2017KUL}, as the LoI would exceed the dimensioning incident and could cause frequency events in the whole synchronous area connected to it, leading to load shedding or generation curtailment. As a result, (partially) selective HVDC protection is likely to be required in EEHs~\cite{leterme2015classification,Leterme2019,SuperGridInstitute}. This kind of protection system relies on DCCBs that can rapidly interrupt DC fault currents and thus prevent the outage of healthy elements in the system. 


Similarly, the control systems of all devices in or connected to the EEH are required to work reliably at any point in time. Malfunctioning controls could lead to the entire trip of the EEH and the subsequent LoI exceeding the dimensioning incident of the power system. Recorded power system events have shown that converter control systems can respond erroneously during faults and frequency excursions, and consequently lead to a subsequent loss of generation and transmission capacity~\cite{NG_Report_2019,NERC_Blue_Cut,NERC_900,AEMO_report}. 
\\

\textbf{\emph{High DC fault currents:}}
The concentration of equipment in EEHs may increase the magnitude of DC-side fault currents. Since several HVDC cables and converters are likely to be connected close to each other, DC faults in or in proximity of the EEH will experience high fault currents that rapidly increase due to the capacitive discharge of these components~\cite{Mudar_DCCB_sizing, Bucher_Franck_capacitive_Fault_currents}. As a result, additional fault current limiting reactors may be required in EEHs to ensure that fault currents do not exceed the maximum current limits of DCCBs. Moreover, given the GW-size power ratings of EEHs, significant energy dissipation equipment may be required in EEHs to dissipate the energy contained in these fault currents~\cite{Mudar_DCCB_sizing}. Opportunities to centralize this energy dissipation equipment may arise depending on the size and topology of the EEH~\cite{Dirk_Multiport_DCCB}. However, it is expected that both the fault current limiting reactors and energy dissipation equipment will significantly contribute to the protection system costs~\cite{jahn2021holistic} (and to the system (in)stability for the reactors~\cite{Thomas_Stability}) and should thus be minimized while considering the constraint of high fault currents in EEHs.\\

\textbf{\emph{Fault detection:}} 
To guarantee the correct functioning of the HVDC protection system, protection algorithms must detect faults in a limited time, while being secure, i.e. not tripping for faults outside of the protected area, as well as sensitive and dependable, i.e. detecting and tripping faults within the protected area~\cite{anderson2022power}. Since the transient fault phenomena in HVDC grids generally happen on a smaller time scale than in AC grids, specific algorithms, such as traveling wave methods, have been developed for HVDC protection~\cite{Xiang_Protection_algorithms}. Moreover, it has been established that in typical multiterminal HVDC grids, algorithms that rely on communication are not applicable due to the long communication delays, especially over long cable distances~\cite{ClausBakAlgorithms}.

However, since EEHs centralize several cables and converters in one location, additional research into the applicability of several algorithms would be required. These studies should evaluate whether typical algorithms are still able to provide the required selectivity given the density of equipment and connections. 

The concentration of equipment in one location may also create opportunities for centralized control, as well as faster communication, thus also creating new opportunities for HVDC fault detection that can be investigated for EEHs. \\

\textbf{\emph{Common mode failures:}}
The high density of electrical equipment in an EEH leads to an increased risk for common mode failures, i.e. the failure of multiple systems due to a common cause. Such events can occur in any system due to the malfunctioning of the protection system, e.g. a circuit breaker failure, or by external causes, such as extreme weather events, fire, terrorist attacks, cyber-attacks, etc.~\cite{Reliability_book}. Since EEHs centralize key grid infrastructure in one location - likely with a high density of equipment due to space limitations - it is expected that they are more prone to common mode failures due to externalities.  Since the loss of the entire EEH could cause a multi-GW LoI with the possibility of a significant impact on connected AC power systems, specific countermeasures have to be taken into the design to avoid such events. Consequently, system operators may opt to operate certain connections as \textit{normally open}, thus creating physical separations to prevent contingencies from impacting the grid at several locations~\cite{Preventive_Decoupling_paper,Elia_Consultation_report}. 
In this sense, System Integrity Protection Schemes (SIPS) are usually established to mitigate large system-wide disturbances caused by faults in the power system. Differently from conventional protection systems related to a specific network element, SIPS provides system-wide countermeasures to slow and/or stop cascading outages caused by extreme contingencies~\cite{SIPS}. Given that faults in EEHs might lead to large imbalances in the power grids to which they are connected, more dynamic and adaptable SIPS are to be developed. They will be particularly needed to prevent cascading events originating from faults in one of the EEH’s network elements.\\

\textbf{\emph{Control interactions:}}
Control interactions refer to the dynamic interplay of controls and components within power systems. Such interactions can take place between the control systems of converters and components, such as cables, overhead lines or generators. 
Given the expected multi-GW size and equipment density of EEHs, control interactions are likely to occur 
if proper design considerations are not taken into account. 
Such control interactions have been observed in practice for offshore wind farms~\cite{buchhagen_borwin1_2015} and HVDC links~\cite{zou_analysis_2018,ji_dc_2021,cigre_Infele} and have been reviewed and summarized in~\cite{Philippe_overview_of_real_life_events}. 
These interactions can cause oscillations across a broad frequency spectrum, from less than 1~Hz to several kHz~\cite{xie_guest_2024}. The oscillations are inherently undesirable as they can damage electrical equipment or cause units to trip, potentially leading to a system collapse. 
Various tools exist to derisk the system from undesired control interactions, such as EMT simulations, RMS simulations, and small-signal analysis~\cite{Cigre_Interactions}. 
These tools allow the proper control design of all units involved in the power aggregation and transmission process of an EEH.

Furthermore, new control functions may be required for EEHs to provide ancillary services, such as oscillation damping \cite{Zhao_forced_Osc}, voltage and frequency support, grid forming \cite{Mian_Cigre}, and black-start capability of the offshore power grid \cite{ Jain_2021}. 
However, these new control functions might lead to previously unexplored interactions, potentially jeopardizing the overall operation of EEHs.\\

\textbf{\emph{Interoperability:}} Interoperability implies that the electrical equipment, e.g. converter stations in MTDC grids, is mutually compatible and interoperable under varying operational modes~\cite{InterOpera.D4, wang2021multi}. As EEHs are centralized multi-GW connection points for HVDC interconnections and RES aggregation, it is expected that further elements could be added in a phased approach, and these elements should be interoperable~\cite{akhmatov2013technical,BriffPESMag2024}. Systems today, although containing components from multiple vendors, are developed by one system integrator (typically a manufacturer). As internal functional requirements and interfaces are commonly set in a proprietary or bespoke way, the viability of future expansion by other parties is inherently limited. Moving beyond these turnkey HVDC systems towards systems for which it is feasible for other manufacturers to add elements in a phased manner is therefore essential for large-scale EEHs. 

Moreover, addressing interoperability issues implies a need for interoperability on several levels~\cite{wang2021multi,PletCigre2022}. Functional interoperability is needed such that devices function correctly together in a system (e.g. converter controls do not adversely interact)~\cite{GeraintCigre}. In fact, the physical electrical interface is required to be interoperable, e.g. in voltage, current and power ratings~\cite{wang2021multi}. Furthermore, for effective system design and operation, manufacturer intellectual property concerns must be mitigated to share key information about systems and components for system design~\cite{Ilka2020}, as well as for signal interfaces~\cite{jahn2022architecture}. Communication interfaces must be well defined with common (standardized) protocols~\cite{akhmatov2013technical}. 
At a fundamental level, system topology should be harmonized, for example which secondary system devices should execute control and protection functions, naturally a precursor to standardized requirements, interfaces and components. 
Addressing these issues is important to ensure a flexible and modular approach to the design and operation of EEHs. Without agreements on these key interoperability topics the design process is challenging due to both technical reasons, e.g. difficulties in predicting and ensuring the stable integration of components without experiencing interactions or incompatibilities, as well as organizational. 




Finally, it is expected that interoperability and viable system expansion can be achieved in future multivendor systems through effective grid codes and standardization~\cite{wang2021multi}. In this manner, a new converter, line, protection IED, or circuit breaker could be added without re-designing the whole system or requiring the involvement of the original manufacturer of other grid elements. However, the industry is still working towards pre-standardization in many fields~\cite{BriffPESMag2024}.
Challenges in multi-party development could be mitigated through future harmonized systems engineering methodologies (e.g. MBSE)~\cite{GeraintCigre}. It is expected that system operators will increasingly add expandability into the functional requirements of new systems~\cite{akhmatov2013technical}, which could be effective as a stepping stone to full HVDC grid codes. 


Given the expectations for EEHs' large-scale energy integration and phased expandability, they are key areas in the HVDC system where multivendor interoperability will be needed. Future EEHs would be expected to be challenged by all of the mentioned interoperability issues, without suitable mitigation through technical solutions governed by grid codes, harmonization and standardization.
\\


\textbf{\emph{Expandability requirements:}} As the aggregation of RES and the interconnections to different control areas are expected to keep increasing in a more electrified future, EEHs are planned to be expandable~\cite{MarsMission}. 
To facilitate such expansions, expansion pathways should be developed in early design stages. 
These pathways give an indication of changes in functional requirements that occur during expansion. For example, to allow the additional cables to be connected to an EEH, besides providing adequate substation space, the system developer should also consider the impact this expansion would have on e.g. the protection strategy requirements. Such changes in protection requirements further impact the substation lay-out, as additional DCCBs and reactors would possibly be required, but also may have implications for the converter fault-ride-through capabilities~\cite{MerijnCigre}. 
These changes are expected to be challenging to implement, especially offshore, given that EEHs are likely to have a high concentration of equipment and are consequently not suitable for the replacement and redesign of existing components. 
Therefore, expansions should be foreseen during the original planning stage of the project, to guarantee that the additional equipment can be safely installed and operated within the EEH context.

\begin{table*}[h!]
\caption{Relation of the identified technical challenges with EEH functionalities.
}
\label{tab:constraints}
\fontsize{8pt}{8.5pt}\selectfont
 \begin{tabular*}{\textwidth}{@{}llc@{}}
\toprule
Technical challenge & Related functionalities & Main Research area\\
\toprule
           & Aggregation & \\
Grid congestion      & Interconnection & Planning\\
 & Multi-GW & \\
\hline
 &  & Control \\
Limit maximum loss of infeed  & \multirow{-2}{*}{Aggregation} & Planning   \\
 & \multirow{-2}{*}{Multi-GW} & Protection \\ 
\hline
        & Multi-GW & \\
High DC fault currents & High density of electrical equipment& Protection \\
        & One unit  &\\
\hline
     &  &  \\
    Fault detection     & \multirow{-2}{*}{High density of electrical equipment} & Protection \\
     & \multirow{-2}{*}{One unit} & \\
 \hline
     & High density of electrical equipment & \\
 Common mode failure     & Interconnection & \multirow{-2}{*}{Control}\\
     & One unit & \multirow{-2}{*}{Protection}\\
 \hline
  & High density of electrical equipment & \\
 Control interactions & Interconnection & Control \\
  & One unit  & \\
  \hline
    &  & \\
Interoperability   & \multirow{-2}{*}{High density of electrical equipment} & \multirow{-2}{*}{Control} \\
  & \multirow{-2}{*}{One unit} & \multirow{-2}{*}{Protection} \\
 \hline
  & Aggregation & \\
   & High density of electrical equipment & \multirow{-2}{*}{Control}\\
  \multirow{-2}{*}{Expandability requirements} & Interconnection & \multirow{-2}{*}{Planning} \\
    & Multifunctional & \multirow{-2}{*}{Protection} \\
 \hline
 \toprule
\end{tabular*}
 \label{tab:challenges}
\end{table*}
\bigbreak
Note that while discussing the regulatory and market design aspects of EEHs is beyond the scope of the paper, it is worth mentioning that recent work~\cite{Hardy, Tosatto, MichielKenisPhD, ELIA_OBZ, TenneT_OBZ, Energinet_OBZ, EU_OBZ} have discussed the impact of different market designs, e.g. home market, single offshore bidding zone, etc., on the overall EEHs' socio-economic welfare. All the above-mentioned references advocate for an Offshore Bidding Zone market design.
While the optimal market design might be case-dependent, its general goal is that the investment cost for society should be minimized while guaranteeing that the owners of the generation assets are remunerated for their investments and the risk associated to them. 

\section{Classification of existing and planned projects}\label{sec:classification}

\begin{table*}[h!]
    \centering
    \caption{Assessment of electrical energy hubs among several grid expansion projects}
    \fontsize{6.6pt}{8.5pt}\selectfont
 \begin{tabular*}{\textwidth}{@{}clc@{}}
 \toprule
     Name & Characteristics & EEH? \\ 
     \toprule
     & 2.262 GW offshore wind  &  \\
     \multirow{-2.0}{*}{Elia's Modular} & AC connections to Belgium & \ding{55} \\
     \multirow{-2.0}{*}{Offshore Grid 1 \cite{Elia_MOG_1}} & No interconnections with multiple control areas  & \\
     \hline
     &  3.5 GW offshore wind &  \\
     Princess Elisabeth & AC connections to Belgium &  \\
     Island~\cite{Eliaenergyisland}  & HVDC interconnections to Belgium, UK, and a Northern country   & \multirow{-2.0}{*}{\ding{51}}\\
     & Located on an artificial island & \\
     \hline
    & Multi-GW offshore wind &  \\
     \multirow{-2.0}{*}{Danish North Sea} & HVDC interconnections to Denmark and other European countries & \ding{51} \\
     \multirow{-2.0}{*}{energy island~\cite{Danish_Energy_Island_NS}} & Located on an artificial island & \\
     \hline
     & 
     3 GW offshore wind & \\
      \multirow{-2.0}{*}{Bornholm energy}  & HVDC interconnections to Denmark and Germany & \ding{51} \\
     \multirow{-2.0}{*}{island~\cite{Bornholm}} & Located on a natural island & \\
    \hline
     & Hub-and-spoke modular concept & \\
     \multirow{-2.0}{*}{North Sea Wind} & Multiple country efforts & \ding{51} \\
    \multirow{-2.0}{*}{Power Hub~\cite{NSWPH}} & Interconnections within different hubs & \\
    \hline
    & 
     Multi-GW offshore wind & \\
     HeideHub~\cite{Heide} & Multiple HVDC interconnections & \ding{55} \\
     & Aggregation and interconnection are not in the same node & \\
    \hline
     &  443 MW onshore wind & \\
     \multirow{-2.0}{*}{Shetland} & First MTDC grid in Europe & \ding{55} \\
     \multirow{-2.0}{*}{MTDC~\cite{Shetland}} & Aggregation and interconnection are not in the same node & \\
     \hline
     & Multi-GW offshore wind &  \\
     StromNetzDC~\cite{Stromnetz} & Multiple HVDC interconnections & \ding{55} \\
     & Aggregation and interconnection are not in the same node & \\
     \hline
          & Multi-GW offshore wind &  \\
     Netherton Hub~\cite{Netherton} & Multiple HVDC interconnections & \ding{55} \\
     & Aggregation and interconnection are not in the same node & \\
     \bottomrule
    \end{tabular*}
    \label{tab:examples}
\end{table*}
Table \ref{tab:examples} presents examples of currently planned projects aggregating bulk generation capacity and transmitting it to main control areas. Given the fact that they combine the two main purposes (\textit{aggregation} and \textit{interconnection}) of EEHs mentioned in Section~\ref{sec:definition}, existing and future projects are evaluated according to the proposed definition to assess whether they can be categorized as an EEH. Four projects, the Princess Elisabeth Island~\cite{Eliaenergyisland}, Danish North Sea energy island~\cite{Danish_Energy_Island_NS}, Bornholm Energy Island~\cite{Bornholm} and North Sea Wind Power Hub~\cite{NSWPH}, fall within the proposed definition. They all aggregate several GW of power in one node with a high density of electrical equipment and are connected to main control areas via several HVDC interconnections. 

In Table~\ref{tab:examples}, six selected projects do not classify as EEHs. Firstly, Elia's Modular Offshore Grid 1~\cite{Elia_MOG_1} aggregates more than 2 GW of offshore wind but does not have several interconnections to main control areas. Secondly, the HeideHub is an onshore substation where the power from two offshore wind farms (LanWin 2~\cite{Lanwin2} and LanWin 3~\cite{Lanwin3}) are collected and transmitted through HVDC links. The wind power is aggregated and converted at several separate offshore converter stations and transferred to the HeideHub through long-distance transmission cables. Therefore, no power is aggregated at the hub itself. Thirdly, the Shetland MTDC grid~\cite{Shetland} is similar to HeideHub, as the wind power has to be transferred with HVDC transmission lines to the node where the multiple interconnections are connected. In addition, the installed wind capacity is lower than the multi-GW functionality discussed in Section~\ref{sec:definition}. Fourthly, the StromNetzDC~\cite{Stromnetz}, and Netherton hub~\cite{Netherton}  projects all host multiple interconnections, but the multi-GW offshore wind RES capacities are not aggregated in the project's location.

Throughout the section, the proposed EEH definition has been utilized to define whether some example projects fall within the EEH category or not. The same classification can be used with future projects when clear distinctions are needed. Examples are the definition of future planning procedures, grid codes, regulations, and support schemes for EEHs and MTDC grids in general, such as the approach currently used by the European Commission for Projects of Common Interest (PCIs)~\cite{PCI_EU}. As with PCIs, the planning and commissioning of projects selected using the proposed EEH definition could be accelerated, guaranteeing the highest benefit to the socioeconomic welfare.

\section{Conclusion}\label{sec:Conclusion}
The offshore grid development plans in Europe combine the aggregation of multi-GW (renewable) power generation capacity and its transmission via interconnections to existing control areas. Since there has not been a comprehensive classification of these projects yet, this paper introduced and defined the term \textit{Electrical Energy Hub}~(EEH). The definition was compared to the (few) existing definitions describing similar projects. It was shown how the proposed terminology and definition describe ongoing grid development projects effectively from a technical perspective, while the limitations of existing terms such as \textit{energy hubs}, \textit{energy islands}, and \textit{offshore energy hub} were discussed. 

By characterizing EEHs from an electrical engineering perspective, additional research needs were identified in key challenge areas such as grid congestion, limit of the maximum loss of infeed, high DC fault currents, fault detection, common mode failure, control interactions, interoperability, and expandability requirements. These challenges were directly related to identified EEH functionalities and assigned to three different research areas, namely control, planning, and protection.

Finally, as an example, the proposed definition was used to classify and characterize well-known existing and planned grid development projects, indicating that there are several ongoing EEH projects, but not all grid expansion projects should be classified as EEHs. 

Thus, this paper consists of a first step to aid system developers with the development of EEHs by identifying and discussing the challenges related to these projects.
It proposes the term EEH as a classification to identify common electrical challenges and solutions related to relevant grid expansion projects.

\bibliographystyle{ieeetr}
\bibliography{References}

\begin{thebibliography}{100}

\bibitem{PV_report}
{International Energy Agency}, ``Renewables 2023 \- {A}nalysis and forecast to 2028.'' [Online] Available: https://www.iea.org/reports/renewables-2023. (accessed March 8, 2024).

\bibitem{Wind_report}
{Global Wind Energy Council}, ``Global wind report 2023.'' [Online] Available: https://gwec.net/globalwindreport2023/. (accessed March 8, 2024).

\bibitem{RenewableEnergyDirective}
{European Parliament} and {Council of the European Union}, ``{Directive (EU) 2023/2413 Of the European Parliament and of the Council of 18 October 2023 amending Directive (EU) 2018/2001, Regulation (EU) 2018/1999 and Directive 98/70/EC as regards the promotion of energy from renewable sources, and repealing Council Directive (EU) 2015/652}.'' [Online] Available: https://eur-lex.europa.eu. (accessed March 27, 2025), 2023.

\bibitem{fitfor55}
{Council of the European Union}, ``Renewable energy directive, fit for 55 packages.'' [Online] Available: https://www.consilium.europa.eu/en/policies/green-deal. (accessed March 8, 2024).

\bibitem{Ostend_declaration}
{Governments of the countries in the North Sea area}, ``Ostend declaration of energy ministers on {T}he {N}orth {S}eas as {E}urope's green power plant delivering cross-border projects and anchoring the renewable offshore industry in {E}urope,'' 24 April 2023.

\bibitem{Elia_MOG_1}
{Elia}, ``{High voltage off the Belgian coast}.'' [Online] Available: https://www.eliagroup.eu/en/publications. (accessed February 21, 2024).

\bibitem{BorWin1}
{TenneT}, ``Borwin 1.'' [Online] Available: https://www.tennet.eu/projects/borwin1. (accessed November 29, 2024).

\bibitem{Dirk2010DCgrids}
D.~Van~Hertem and M.~Ghandhari, ``{Multi-terminal VSC HVDC for the European supergrid: Obstacles},'' {\em Renewable and Sustainable Energy Reviews}, vol.~14, no.~9, pp.~3156--3163, 2010.

\bibitem{IEA_grid}
{International Energy Agency}, ``Electricity grids and secure energy transitions.'' [Online] Available: https://www.iea.org/reports/electricity-grids-and-secure-energy-transitions. (accessed March 8, 2024).

\bibitem{Grid_action_plan}
{European Commission}, ``Grids, the missing link - {A}n {EU} action plan for grids.'' [Online] Available: https://eur-lex.europa.eu. (accessed March 8, 2024).

\bibitem{ONDP}
ENTSO-E, ``European offshore network transmission infrastructure needs.'' [Online] Available: https://www.entsoe.eu/outlooks/offshore-hub/tyndp-ondp/. (accessed March 8, 2024).

\bibitem{Eliaenergyisland}
{Elia Group}, ``Transforming our seas into {E}urope’s sustainable economic engine.'' [Online] Available: https://www.eliagroup.eu/en/publications. (accessed November 23, 2023).

\bibitem{Bornholm}
{Danish Energy Agency}, ``Bornholm energy island.'' [Online] Available: https://ens.dk/en/our-responsibilities/onshore-wind-power/bornholm-energy-island. (accessed February 14, 2024).

\bibitem{Danish_Energy_Island_NS}
{Danish Energy Agency}, ``Energy {I}sland in the {N}orth {S}ea.'' [Online] Available: https://ens.dk/en/our-responsibilities/offshore-wind-power/energy-island-north-sea. (accessed March 4, 2024).

\bibitem{Tennet_2GW}
{TenneT}, ``2 {GW} program.'' [Online] Available: https://www.tennet.eu/about-tennet/innovations/2gw-program. (accessed March 8, 2024).

\bibitem{bookDirk}
D.~Van~Hertem, O.~Bellmunt, and J.~Liang, {\em {HVDC} grids for transmission of electrical energy: {O}ffshore grids and a future supergrid}.
\newblock Wiley, 2016.

\bibitem{Caith_Moray}
{Hitachi Energy}, ``{Caithness {M}oray {HVDC} {L}ink}.'' [Online] Available: https://www.hitachienergy.com/news-and-events/customer-success-stories/caithness-moray-hvdc-link. (accessed September 13, 2024), {2024}.

\bibitem{DolWin2}
{TenneT}, ``Dolwin 2.'' [Online] Available: https://www.tennet.eu/projects/dolwin2. (accessed November 29, 2024).

\bibitem{DolWin3}
{TenneT}, ``Dolwin 3.'' [Online] Available: https://www.tennet.eu/projects/dolwin3. (accessed November 29, 2024).

\bibitem{HollandseKust}
{Vattenfall}, ``{Hollandse Kust Zuid Wind Farm}.'' [Online] Available: https://hollandsekust.vattenfall.nl/en/wind-farm/. (accessed October 22, 2024).

\bibitem{BorWin2}
{TenneT}, ``Borwin 2.'' [Online] Available: https://www.tennet.eu/projects/borwin2. (accessed November 29, 2024).

\bibitem{BorWin3}
{TenneT}, ``Borwin 3.'' [Online] Available: https://www.tennet.eu/projects/borwin3. (accessed November 29, 2024).

\bibitem{DolWin6}
{TenneT}, ``Dolwin 6.'' [Online] Available: https://www.tennet.eu/projects/dolwin6. (accessed November 29, 2024).

\bibitem{Hornsea1}
{Orsted}, ``{Hornsea 1 Wind Farm}.'' [Online] Available: https://orsted.co.uk/energy-solutions/offshore-wind/our-wind-farms/hornsea1. (accessed October 22, 2024).

\bibitem{Stephen_powertech}
S.~Hardy, D.~Van~Hertem, and H.~Ergun, ``A techno-economic analysis of meshed topologies of offshore wind hvac transmission,'' in {\em 2021 IEEE Madrid PowerTech}, pp.~1--6, 2021.

\bibitem{Meijden_Tennet}
M.~van~der Meijden, ``Future north sea infrastructure based on dogger bank modular island,'' in {\em 15th Wind Integration Workshop (WIW), Vienna, Austria}, 2016.

\bibitem{NSWPH}
{North Sea Wind Power Hub}, ``Key concepts for realizing the potential of offshore wind.'' [Online] Available: https://northseawindpowerhub.eu/key-concepts. (accessed February 14, 2024).

\bibitem{PROMOTioN_final_report}
{PROMOTioN Project}, ``{Final report}.'' [Online] Available: https://www.promotion\-offshore.net. (accessed February 21, 2024), 2020.

\bibitem{Shetland}
{SSEN Transmission}, ``{Shetland HVDC Link}.'' [Online] Available: https://www.ssen-transmission.co.uk/projects/project-map/shetland/. (accessed September 15, 2024).

\bibitem{MarsMission}
N.~Cutululis, F.~Blaabjerg, J.~{\O}stergaard, C.~Bak, M.~Anderson, F.~Silva, H.~Johannsson, X.~Wang, and B.~J{\o}rgensen, ``{The Energy Islands: A Mars Mission for the Energy System},'' 2021.

\bibitem{PCI}
{European Commission and European Climate, Infrastructure and Environment Executive Agency}, {\em Connecting {E}urope {F}acility – {E}nergy – {S}upported actions 2014-2020 – {U}pdate {M}ay 2021}.
\newblock Publications Office of the European Union, 2021.

\bibitem{ENTSO_E_CBA}
ENTSO-E, ``4th {ENTSO-E} {G}uideline for {C}ost {B}enefit {A}nalysis of {G}rid {D}evelopment {P}rojects.'' [Online] Available: https://eepublicdownloads.blob.core.windows.net/public-cdn-container/tyndp-documents/CBA/CBA4/221215\_CBA4-Guideline\_v1.0\_for-public-consultation.pdf. (accessed February 14, 2024).

\bibitem{Glossary}
{Union for Co-ordination of Transmission of Electricity}, ``{UCTE Operation Handbook - Glossary}.'' [Online] Available: https://www.ucte.org/\_library/ohb/glossary\_v22.pdf. (accessed February 4, 2025), 2019.

\bibitem{geidl2006energy}
M.~Geidl, G.~Koeppel, P.~Favre-Perrod, B.~Klockl, G.~Andersson, and K.~Frohlich, ``Energy hubs for the future,'' {\em IEEE power and energy magazine}, vol.~5, no.~1, pp.~24--30, 2006.

\bibitem{LUTH_H2}
A.~Lüth, Y.~Werner, R.~Egging-Bratseth, and J.~Kazempour, ``Electrolysis as a flexibility resource on energy islands: {T}he case of the {N}orth {S}ea,'' {\em Energy Policy}, vol.~185, p.~113921, 2024.

\bibitem{LUTH_2024}
A.~Lüth and D.~Keles, ``Risks, strategies, and benefits of offshore energy hubs: A literature-based survey,'' {\em Renewable and Sustainable Energy Reviews}, vol.~203, p.~114761, 2024.

\bibitem{capacity_and_congestion}
{European Union}, ``{Commission Regulation (EU) 2015/1222 of 24 July 2015: {E}stablishing a guideline on capacity allocation and congestion management (Text with EEA relevance)}.'' [Online] Available: https://eur-lex.europa.eu/eli/reg/2015/1222/oj/eng (accessed November 19, 2024), {2015}.

\bibitem{Hakan_powertech}
H.~Ergun, B.~Rawn, R.~Belmans, and D.~Van~Hertem, ``Optimization of transmission technology and routes for pan-european electricity highways considering spatial aspects,'' in {\em 2015 IEEE Eindhoven PowerTech}, pp.~1--6, 2015.

\bibitem{PSTs_placement}
M.~Franken, H.~Barrios, A.~B. Schrief, and A.~Moser, ``Transmission expansion planning via power flow controlling technologies,'' {\em IET Generation, Transmission \& Distribution}, vol.~14, no.~17, pp.~3530--3538, 2020.

\bibitem{Morsy2022}
B.~Morsy, A.~Hinneck, D.~Pozo, and J.~Bialek, ``Security constrained {OPF} utilizing substation reconfiguration and busbar splitting,'' {\em Electr. Power Syst. Res.}, vol.~212, p.~108507, Nov. 2022.

\bibitem{Hinneck}
A.~Hinneck, B.~Morsy, D.~Pozo, and J.~Bialek, ``Optimal power flow with substation reconfiguration,'' in {\em 2021 IEEE Madrid PowerTech}, pp.~1--6, 2021.

\bibitem{Heidarifar2016}
M.~Heidarifar and H.~Ghasemi, ``A network topology optimization model based on substation and node-breaker modeling,'' {\em IEEE Transactions on Power Systems}, vol.~31, no.~1, pp.~247--255, 2016.

\bibitem{nrao}
{CCR Core TSO Cooperation}, ``Explanatory document to the \uppercase{C}ore \uppercase{C}apacity \uppercase{C}alculation \uppercase{R}egion methodology for common provisions for regional operational security coordination in accordance with \uppercase{A}rticle 76 of \uppercase{C}ommission \uppercase{R}egulation (\uppercase{E}\uppercase{U}) 2017/1485,'' 2017.

\bibitem{Jack}
G.~Bastianel, M.~Vanin, D.~{Van Hertem}, and H.~Ergun, ``{Optimal Transmission Switching and Busbar Splitting in Hybrid AC/DC Grids}.'' arXiv:2412.00270, 2024.

\bibitem{Commission_regulation}
{European Union}, ``{Commission Regulation (EU) No 543/2013 of 14 June 2013 on submission and publication of data in electricity markets and amending Annex I to Regulation (EC) No 714/2009 of the European Parliament and of the Council}.'' [Online] Available: https://eur-lex.europa.eu (Accessed November 19, 2024), {2013}.

\bibitem{ENTSO_E_Congestion}
{ENTSO-e Transparency Platform}, ``{Costs of Congestion Management}.'' [Online] Available: https://transparency.entsoe.eu/congestion-management/r2/costs/show. (Accessed November 27, 2024).

\bibitem{EC_grid_code}
{European Commission}, ``Commission {R}egulation ({EU}) 2017/1485 of 2 august 2017 establishing a guideline on electricity transmission system operation.'' [Online] Available: https://eur-lex.europa.eu. (accessed September 24, 2024).

\bibitem{Elia2023Website}
{Elia Group}, ``Princess {E}lisabeth {I}sland.'' [Online] Available: https://www.elia.be/en/infrastructure-and-projects/infrastructure-projects/princess-elisabeth-island. (accessed November 17, 2023).

\bibitem{DCcont2017KUL}
M.~Abedrabbo, M.~Wang, P.~Tielens, F.~Z. Dejene, W.~Leterme, J.~Beerten, and D.~{Van Hertem}, ``{Impact of DC grid contingencies on AC system stability},'' in {\em Proc. IET ACDC 2017}, (Manchester, UK), 2017.

\bibitem{leterme2015classification}
W.~Leterme and D.~Van~Hertem, ``{Classification of Fault Clearing Strategies for HVDC Grids},'' in {\em CIGRE, Lund}, 2015.

\bibitem{Leterme2019}
W.~Leterme, I.~Jahn, P.~Ruffing, K.~Sharifabadi, and D.~V. Hertem, ``{Designing for High-Voltage dc Grid Protection: Fault Clearing Strategies and Protection Algorithms},'' {\em IEEE Power and Energy Magazine}, vol.~17, pp.~73--81, 5 2019.

\bibitem{SuperGridInstitute}
{SuperGrid Institute}, ``{NSWPH Validation Technical Requirements: SoW A Final Feasibility Report}.'' [Online] Available: https://northseawindpowerhub.eu/knowledge/nswph-validation-technical-requirements (accessed December 12, 2024), 2022.

\bibitem{NG_Report_2019}
{National Grid}, ``{Technical report on the events of 9 August 2019}.'' [Online] Available: https://www.nationalgrideso.com. (accessed August 20, 2024).

\bibitem{NERC_Blue_Cut}
{NERC}, ``{1200 {MW} Fault Induced Solar Photovoltaic Resource Interruption Disturbance Report}.'' [Online] Available: https://www.nerc.com/pa/rrm/ea/1200{\_}MW{\_}Fault{\_}Induced{\_}Solar{\_} \sloppy Photovoltaic{\_}Resource{\_}/1200{\_}MW{\_}Fault{\_}Induced{\_}Solar{\_} \sloppy Photovoltaic{\_}Resource {\_}Interruption{\_}Final.pdf (accessed October 10, 2024).

\bibitem{NERC_900}
{NERC}, ``{900 {MW} Fault Induced Solar Photovoltaic Resource Interruption Disturbance Report}.'' [Online] Available: https://www.nerc.com/pa/rrm/ea/october{\%}209{\%}202017{\%}20canyon \sloppy {\%}202{\%}20fire{\%}20disturbance{\%}20report/900{\%}20mw{\%}20solar{\%}20 \sloppy photovoltaic{\%}20resource{\%}20interruption{\%}20disturbance{\%}20report \sloppy .pdf (accessed October 15, 2024).

\bibitem{AEMO_report}
{Australian Energy Market Operator}, ``{Black System South Australia 28 September 2016}.'' [Online] Available: https://apo.org.au/sites/default/files/resource-files/2017-03/apo-nid74886.pdf (accessed October 15, 2024).

\bibitem{Mudar_DCCB_sizing}
M.~Abedrabbo, W.~Leterme, and D.~{Van Hertem}, ``{Systemic Approach to HVDC Circuit Breaker Sizing},'' {\em IEEE Transactions on Power Delivery}, vol.~35, no.~1, pp.~288--300, 2019.

\bibitem{Bucher_Franck_capacitive_Fault_currents}
M.~K. Bucher and C.~M. Franck, ``Analytic approximation of fault current contributions from capacitive components in hvdc cable networks,'' {\em IEEE Transactions on Power Delivery}, vol.~30, no.~1, pp.~74--81, 2015.

\bibitem{Dirk_Multiport_DCCB}
A.~Mokhberdoran, D.~Van~Hertem, N.~Silva, H.~Leite, and A.~Carvalho, ``Multiport hybrid hvdc circuit breaker,'' {\em IEEE Transactions on Industrial Electronics}, vol.~65, no.~1, pp.~309--320, 2017.

\bibitem{jahn2021holistic}
I.~Jahn, G.~Chaffey, N.~Svensson, and S.~Norrga, ``A holistic method for optimal design of hvdc grid protection,'' {\em Electric Power Systems Research}, vol.~196, p.~107234, 2021.

\bibitem{Thomas_Stability}
T.~Roose, A.~Lekić, M.~M. Alam, and J.~Beerten, ``{Stability Analysis of High-Frequency Interactions Between a Converter and HVDC Grid Resonances},'' {\em IEEE Transactions on Power Delivery}, vol.~36, no.~6, pp.~3414--3425, 2021.

\bibitem{anderson2022power}
P.~M. Anderson, C.~F. Henville, R.~Rifaat, B.~Johnson, and S.~Meliopoulos, {\em Power system protection}.
\newblock John Wiley \& Sons, 2022.

\bibitem{Xiang_Protection_algorithms}
W.~Xiang, S.~Yang, G.~P. Adam, H.~Zhang, W.~Zuo, and J.~Wen, ``{DC Fault Protection Algorithms of MMC-HVDC Grids: Fault Analysis, Methodologies, Experimental Validations, and Future Trends},'' {\em IEEE Transactions on Power Electronics}, vol.~36, no.~10, pp.~11245--11264, 2021.

\bibitem{ClausBakAlgorithms}
M.~Ashouri, C.~L. Bak, and F.~Faria Da~Silva, ``{A review of the protection algorithms for multi-terminal VCD-HVDC grids},'' in {\em 2018 IEEE International Conference on Industrial Technology (ICIT)}, pp.~1673--1678, 2018.

\bibitem{Reliability_book}
G.~Macangus-Gerrard, ``Chapter 1 - reliability,'' in {\em Offshore Electrical Engineering Manual (Second Edition)} (G.~Macangus-Gerrard, ed.), pp.~387--406, Boston: Gulf Professional Publishing, second edition~ed., 2018.

\bibitem{Preventive_Decoupling_paper}
P.~Düllmann, C.~Klein, P.~Winter, H.~Köhler, M.~Steglich, J.~Teuwsen, and A.~Moser, ``{Preventive DC-side decoupling: a control and operation concept to limit the impact of DC faults in offshore multi-terminal HVDC systems},'' {\em 19th International Conference on AC and DC Power Transmission (ACDC 2023)}, pp.~30--37, 2023.

\bibitem{Elia_Consultation_report}
{Elia}, ``{Public Consultation Task Force Princess Elisabeth Zone}.'' {[Online] Available: https://www.elia.be/en/public-consultation/20231120\_public-consultation-task-force-princess-elisabeth-zone. (accessed December 2, 2024)}, 2024.

\bibitem{SIPS}
V.~Madani, D.~Novosel, S.~Horowitz, M.~Adamiak, J.~Amantegui, D.~Karlsson, S.~Imai, and A.~Apostolov, ``{IEEE PSRC Report on Global Industry Experiences With System Integrity Protection Schemes (SIPS)},'' {\em IEEE Transactions on Power Delivery}, vol.~25, no.~4, pp.~2143--2155, 2010.

\bibitem{buchhagen_borwin1_2015}
C.~Buchhagen, C.~Rauscher, A.~Menze, and J.~Jung, ``{BorWin}1 - first experiences with harmonic interactions in converter dominated grids,'' in {\em International {ETG} Congress 2015; Die Energiewende - Blueprints for the new energy age}, pp.~1--7, 2015.

\bibitem{zou_analysis_2018}
C.~Zou, H.~Rao, S.~Xu, Y.~Li, W.~Li, J.~Chen, X.~Zhao, Y.~Yang, and B.~Lei, ``Analysis of resonance between a {VSC}-{HVDC} converter and the {AC} grid,'' {\em {IEEE} Transactions on Power Electronics}, vol.~33, no.~12, pp.~10157--10168, 2018.

\bibitem{ji_dc_2021}
K.~Ji, H.~Pang, J.~Yang, and G.~Tang, ``{DC} side harmonic resonance analysis of {MMC}-{HVDC} considering wind farm integration,'' {\em {IEEE} Transactions on Power Delivery}, vol.~36, no.~1, pp.~254--266, 2021.

\bibitem{cigre_Infele}
H.~Saad, Y.~Vernay, S.~Dennetiere, P.~Rault, and B.~Clerc, ``System dynamic studies of power electronics devices with real-time simulation - {A} {TSO} operational experience,'' in {\em CIGRE Session}, CIGRE, 2018.

\bibitem{Philippe_overview_of_real_life_events}
P.~{De Rua}, T.~Roose, Özgür Can~Sakinci, N.~{de Morais Dias Campos}, and J.~Beerten, ``Identification of mechanisms behind converter-related issues in power systems based on an overview of real-life events,'' {\em Renewable and Sustainable Energy Reviews}, vol.~183, p.~113431, 2023.

\bibitem{xie_guest_2024}
X.~Xie, J.~Shair, J.~Beerten, L.~Fan, O.~Gomis-Bellmunt, P.~Vorobev, R.~Preece, S.~Shah, X.~Wang, Y.~Wang, and V.~Terzija, ``Guest editorial: Control interactions in power electronic converter dominated power systems,'' {\em International Journal of Electrical Power \& Energy Systems}, vol.~155, p.~109553, 2024.

\bibitem{Cigre_Interactions}
{CIGRE Working Group B4.81}, ``Interaction between nearby {VSC}-{HVDC} converters, {FACTs} devices, {HV} power electronic devices and conventional {AC} equipment,'' 2024.

\bibitem{Zhao_forced_Osc}
X.~Zhao, Y.~Xue, and X.-P. Zhang, ``Isolation and suppression of forced oscillations through wind farms under grid following and grid forming control,'' {\em IEEE Access}, vol.~9, pp.~76446--76460, 2021.

\bibitem{Mian_Cigre}
M.~Wang, B.~Strong, H.~Bouattour, N.~Cherouvim, and S.~Hug, ``{Grid Forming Capabilities of HVDC Systems Connecting Offshore Wind Parks},'' in {\em CIGRE Session}, 09 2023.

\bibitem{Jain_2021}
A.~Jain, {\em Green \& Black-starting HVDC-connected Offshore Wind Power Plants: Grid forming control, Energization transients and Islanding capabilities}.
\newblock PhD thesis, DTU, Denmark, 2021.

\bibitem{InterOpera.D4}
G.~Dawson and C.~T. Nieuwenhout, ``{Deliverable 4.2 Multi-Part Cooperation Framework-Preliminary draft with focus on information sharing}.'' [Online] Available: https://interopera.eu/publications/ (accessed December 12, 2024), 2023.

\bibitem{wang2021multi}
M.~Wang, W.~Leterme, G.~Chaffey, J.~Beerten, and D.~Van~Hertem, ``{Multi-vendor interoperability in HVDC grid protection: State-of-the-art and challenges ahead},'' {\em IET Generation, Transmission \& Distribution}, vol.~15, no.~15, pp.~2153--2175, 2021.

\bibitem{akhmatov2013technical}
V.~Akhmatov, M.~Callavik, C.~Franck, S.~E. Rye, T.~Ahndorf, M.~K. Bucher, H.~M{\"u}ller, F.~Schettler, and R.~Wiget, ``Technical guidelines and prestandardization work for first hvdc grids,'' {\em IEEE transactions on power delivery}, vol.~29, no.~1, pp.~327--335, 2013.

\bibitem{BriffPESMag2024}
P.~Briff, L.~Zou, H.~Schuldt, F.~Schettler, C.~Wikström, P.~Lundberg, and D.~Kolichev, ``Achieving interoperability for multiterminal multivendor hvdc systems: Exploring the main challenges,'' {\em IEEE Power and Energy Magazine}, vol.~22, no.~5, pp.~49--59, 2024.

\bibitem{PletCigre2022}
C.~Plet, C.~Brantl, M.~Wang, H.~Evans, J.~Moore, C.~Nieuwenhout, A.~Armeni, D.~{van Hertem}, and M.~Semenyuk, ``Compatibility \& interoperability framework to facilitate the step-wise organic development of multi-terminal hvdc grids,'' in {\em Cigre Paris Session}, 2022.

\bibitem{GeraintCigre}
G.~Chaffey, I.~Jahn, M.~Hoffmann, R.~{Alvarez Valenzuela}, E.~Prieto, and S.~Norrga, ``{Model-based systems engineering for HVDC grids - state-of-the-art and future outlook},'' in {\em Cigre Paris Session, 2024}, 2024.

\bibitem{Ilka2020}
I.~Jahn, F.~Hohn, K.~Sharifabadi, M.~Wang, G.~Chaffey, and S.~Norrga, ``{Requirements for open specifications in multivendor HVDC protection systems},'' in {\em 15th International Conference on Developments in Power System Protection (DPSP 2020)}, 2020.

\bibitem{jahn2022architecture}
I.~Jahn, M.~Nahalparvari, C.~Hirsching, M.~Hoffmann, P.~D{\"u}llmann, F.~Loku, A.~Agbemuko, G.~Chaffey, E.~Prieto-Araujo, and S.~Norrga, ``An architecture for a multi-vendor vsc-hvdc station with partially open control and protection,'' {\em IEEE Access}, vol.~10, pp.~13555--13569, 2022.

\bibitem{MerijnCigre}
M.~{Van Deyck}, G.~Chaffey, M.~Abedrabbo, H.~Ergun, D.~{Van Hertem}, E.~Spahic, and D.~{De Decker}, ``{Expansion planning of offshore HVDC grids considering protection system design},'' in {\em Cigre Paris Session, 2024}, 2024.

\bibitem{Hardy}
S.~Hardy, A.~Themelis, K.~Yamamoto, H.~Ergun, and D.~Van~Hertem, ``Optimal grid layouts for hybrid offshore assets in the north sea under different market designs,'' {\em IEEE Transactions on Energy Markets, Policy and Regulation}, vol.~1, no.~4, pp.~468--479, 2023.

\bibitem{Tosatto}
A.~Tosatto, X.~M. Beseler, J.~Østergaard, P.~Pinson, and S.~Chatzivasileiadis, ``{North Sea Energy Islands: Impact on national markets and grids},'' {\em Energy Policy}, vol.~167, p.~112907, 2022.

\bibitem{MichielKenisPhD}
M.~Kenis, {\em {Offshore Wind Power in Electricity Markets, Regulatory and Market Design Implications under Flow-Based Market Coupling}}.
\newblock PhD thesis, KU Leuven, 2023.

\bibitem{ELIA_OBZ}
{Elia}, ``{Workshop {MOG} 2}.'' [Online] Available at: https://www.elia.be/\-/media/project/elia/elia\-site/users\-group/ug/workshop/2023/20230206/20230206\_mog\_2\_market\-design\_elia\_website.pdf (accessed 8 August, 2024), {2023}.

\bibitem{TenneT_OBZ}
{TenneT}, ``{The offshore bidding zone - a blueprint by {T}enne{T}}.'' [Online] Available: https://www.elia.be/-/media/project/elia/elia-site/users-group/ug/workshop/2023/20230206/20230206\_mog\_2\_market-design\_elia\_website.pdf (accessed August 9, 2024), {2024}.

\bibitem{Energinet_OBZ}
{Energinet}, ``{The ideal market design for offshore grids - {A} {N}ordic {TSO} perspective}.'' [Online] Available: https://en.energinet.dk/About\-our\-news/News/2020/11/04/Ideal\-market\-design\-fo\r-ofshore\-grids/ (accessed August 9, 2024), {2020}.

\bibitem{EU_OBZ}
{Engie Impact}, ``{Support on the use of congestion revenues for Offshore Renewable Energy Projects connected to more than one market}.'' [Online] Available: https://www.elia.be/\-/media/project/elia/elia\-site/users-group/ug/workshop/2023/20230206/20230206\_mog\_2\_market-design\_elia\_website.pdf (accessed August 9, 2024), {2022}.

\bibitem{Heide}
Tennet, ``Tennet {O}ffshore.'' [Online] Available: https://www.tennet.eu/offshore-overview\#16655. (accessed January 15, 2024).

\bibitem{Stromnetz}
{50 Hertz, TenneT and Transnet BW}, ``{Zusammen für die Energiewende – StromNetzDC}.'' [Online] Available: https://www.stromnetzdc.com/. (accessed December 10, 2024), 2024.

\bibitem{Netherton}
{Scottish and Southern electricity networks}, ``{Netherton Hub}.'' [Online] Available: https://www.ssen-transmission.co.uk/projects/project-map/netherton-hub/. (accessed December 10, 2024), 2024.

\bibitem{Lanwin2}
{TenneT}, ``{LanWin2}.'' [Online] Available: https://www.tennet.eu/projects/lanwin2. (accessed September 26, 2024), 2024.

\bibitem{Lanwin3}
{50 Hertz}, ``{LanWin3}.'' [Online] Available: www.50hertz.com/en/Grid/Griddevelopement/Offshoreprojects. (accessed September 26, 2024), 2024.

\bibitem{PCI_EU}
{European Union}, ``{Regulation (EU) 2022/869 of the European Parliament and of the Council of 30 May 2022 on guidelines for trans-European energy infrastructure, amending Regulations (EC) No 715/2009, (EU) 2019/942 and (EU) 2019/943 and Directives 2009/73/EC and (EU) 2019/944, and repealing Regulation (EU) No 347/2013}.'' [Online] Available: https://eur-lex.europa.eu/homepage.html?lang\=en (accessed December 10, 2024), {2022}.

\end{thebibliography}

\section*{Acknowledgement}
This paper has received support from the Belgian
Energy Transition Fund, FOD Economy, project DIRECTIONS.


\end{document}